\documentclass[twocolumn,aps,showpacs,nofootinbib,floatfix]{revtex4} 

\usepackage{amsmath} 
\usepackage{graphicx}
\usepackage{dcolumn}
\usepackage{bm}

\begin{document}

\title{Large-scale second RPA calculations with finite-range interactions} 

\author{P.~Papakonstantinou} 
\email[Email:]{panagiota.papakonstantinou@physik.tu-darmstadt.de} 

\author{R.~Roth} 

\affiliation{Institut f\"ur Kernphysik, 
Technische Universit\"at Darmstadt, 
Schlossgartenstr.~9, 
D-64289 Darmstadt, Germany} 

\begin{abstract} 
Second RPA (SRPA) calculations of nuclear response are performed and analyzed. 
Unlike in most other SRPA applications, the ground state, 
approximated by the Hartree-Fock (HF) ground state, 
and the residual couplings are described by the same Hamiltonian and  
no arbitrary truncations are imposed on the model space. 
Finite-range interactions are used and thus divergence problems are not present. 
We employ a realistic interaction, derived from the Argonne V18 potenial 
using the unitary correlation operator method (UCOM), 
as well as the simple Brink-Boeker interaction. 
Representative results are discussed, mainly on giant resonances and 
low-lying collective states. 
The focus of the present work is not on the comparison with data, 
but rather on technical and physical aspects of the method. 
We present 
how the large-scale eigenvalue problem that SRPA entails can be treated, 
and demonstrate 
how the method operates in producing self-energy corrections and fragmentation. 
The so-called diagonal approximation is conditionally validated. 
Stability problems are traced back to missing ground-state correlations. 
\end{abstract} 



\maketitle 

\section{Introduction} 

Extended RPA theories such as second RPA (SRPA)~\cite{SpW1991}, which go beyond first-order RPA, are often used to describe the strength, decay width and fine structure of nuclear giant resonances (GRs) and other collective low-lying states. 
They also appear particularly useful in the case of unitarily transformed realistic interactions, which 
have not been calibrated for first-order RPA methods, but rather rely on extended model spaces to provide 
converged results. 
Such interactions also make ``self-consistent" extended-RPA applications possible, in the sense that 
the ground state and the residual couplings can be described by the same Hamiltonian. 
First applications using a renormalized Argonne V18 potential, 
derived with the unitary correlation operator method (UCOM)~\cite{RNH2004,RHP2005}, were presented in Ref.~\cite{PaR2009}. 

Self-consistent extended-RPA applications in large spaces without arbitrary truncations can be envisioned, in principle, with any properly constructed finite-range effective interaction. 
However, they are hardly ever performed for various technical and conceptual reasons. 
Phenomenological effective interactions are fitted to sets of experimental data using mostly Hartree-Fock(-Bogoliubov) and selected (quasi-particle) RPA results. 
Their range of applicability is then restricted to the selected observables and many-body methods.  
Zero-range effective interactions, which greatly simplify numerical applications, cannot be employed in 
second-order RPA methods, because they do not provide a natural cutoff in momentum space. 
Consequently, no effective interactions have been fitted to calculations beyond first-order RPA 
and consistency in the treatment of the ground and excited states is ordinarily abandoned in practical applications of such methods. 

It is the purpose of the present work to perform and analyze large-scale (i.e., without arbitrary truncations), ``self-consistent" (i.e., with a single interaction as the sole input) SRPA calculations. 
We employ mostly the $V_{\mathrm{UCOM}}$ interaction used in Ref.~\cite{PaR2009}.  
It is derived from the Argonne V18 potential by means of a unitary transformation, which renormalizes it, while preserving the phase shifts and retaining the complex structure of the realistic interaction. 
We also use the 
Brink-Boeker interaction, $V_{\mathrm{BB}}$~\cite{BrB1967},  
which is a simple, central, phenomenological effective interaction. 
No explicit three-body force is used at this point. 
We will not focus on producing realistic results and comparing them with data, as was done in Ref.~\cite{PaR2009}, but rather on technical and physical aspects of the method.  
We present how the large-scale eigenvalue problems that SRPA involves can be treated, demonstrate how the method operates in producing self-energy corrections and fragmentation, and discuss consistency and stability problems. 

In the next section we present the SRPA formalism and in Sec.~\ref{S:solve} the methods we have used to solve the SRPA eigenvalue problem. In Sec.~\ref{S:res} we discuss our results with the help of illustrative examples. We conclude in Sec.~\ref{S:concl}.  

\section{Second RPA formalism} 
\label{S:srpa} 

In the following we assume a nuclear Hamiltonian consisting, 
in general, of a one-body part and a two-body part, 
\[ H = H_1 + H_2 .\] 
Three-body terms are not included. 
If the total Hamiltonian is considered, then 
\[ H_1 = T = \frac{1}{2m}\sum_{i=1}^{A}p_i^2\] 
is the total kinetic energy, while $H_2=V$ contains the interactions of particle pairs. 
If the intrinsic Hamiltonian is considered, then $H_1=0$ and $H_2=T_{\mathrm{int}}+V$ 
includes the intrinsic 
kinetic energy of the system, 
\[ T_{\mathrm{int}}=\frac{1}{2mA}\sum_{i<j}(\vec{p}_i-\vec{p}_j)^2.\] 

We will employ the SRPA as it was formulated in Ref.~\cite{Yan1987} in analogy to RPA. 
The derivation is based on the equations-of-motion method and relies on a quasi-boson approximation.  
We will consider closed-(sub)shell spherical nuclei and their excited states of definite angular momentum and parity $J^{\pi}$. Excited states are expanded in the space of particle-hole ($ph$) and two-particle-two-hole ($2p2h$) configurations.  
The symbol $p$ (or $h$) will represent all the quantum numbers of a particle 
(hole) state except the magnetic quantum number $m_p$ $(m_h)$, 
i.e.,  the set of quantum numbers $\{n_{p(h)} \ell_{p(h)} j_{p(h)} t_{p(h)} \}$ 
of the nodes ($n=0,1,\ldots$), orbital angular momentum, total angular momentum, 
and isospin. 
The combined label $(\ell j)_{\alpha} = j_{\alpha} + \ell_{\alpha} - 1/2$ is used to label the $\ell j $ combination uniquely; then $\ell = [(\ell j + 1)/2 ]$ and $j=[\ell j/2]+1/2$ 
(where $[x]$ is the integer part of $x$).   
The Greek letters $\alpha , \beta , ... $  
will be used to denote single particle states of either kind ($p$ or $h$).  

The operator $Q_{\lambda}^{\dagger}$ that creates an 
excited state $|\lambda\rangle $ of energy $E_{\lambda}=\hbar\omega_{\lambda}$ with respect to the $0^+$ ground state $|0\rangle$ and of angular momentum $JM$, 
\begin{equation} 
|\lambda ; JM \rangle = Q_{\lambda ; JM}^{\dagger} |0\rangle \, , 
\,\,\, 
Q_{\lambda ; JM} |0\rangle = 0 \, , 
\end{equation} 
is written as 
\begin{eqnarray}  
Q_{\lambda ; JM}^{\dagger} &=& 
\sum_{ph} X_{ph}^{\lambda ; JM} {O^{JM^{\dagger}}_{ph}} 
- (-1)^{J+M}\sum_{ph} Y_{ph}^{\lambda ;JM} O^{J\, -M}_{ph} 
\nonumber \\ 
&&  
 + \!\! \sum_{p_1\leq p_2,h_1\leq h_2;J_p,J_h} \!\! 
    \mathcal{X}_{p_1h_1p_2h_2J_pJ_h}^{\lambda ; JM} {\mathcal{O}^{JM^{\dagger}}_{p_1h_1p_2h_2J_pJ_h}} 
\nonumber \\ 
&&  
 - (-1)^{J+M} \!\! \sum_{p_1\leq p_2, h_1\leq h_2} \!\! 
      \mathcal{Y}_{p_1h_1p_2h_2J_pJ_h}^{\lambda ; JM} \mathcal{O}^{J\, -M}_{p_1h_1p_2h_2J_pJ_h} 
, 
\label{E:Qop}  
\end{eqnarray} 
where $O^{JM^{\dagger}}_{ph}$ 
creates a $ph$ state and 
${\mathcal{O}^{JM^{\dagger}}_{p_1h_1p_2h_2J_pJ_h}}$ 
creates a $2p2h$ state, coupled to the given quantum numbers. 
In particular, we have  
\begin{equation} 
O^{JM^{\dagger}}_{ph} = \sum_{m_p,m_h} (-1)^{j_h-m_h}\langle j_p m_p j_h -m_h | JM_J \rangle 
a_{pm_p}^{\dagger} a_{hm_h}   
\end{equation} 
\begin{eqnarray} 
\mathcal{O}^{JM^{\dagger}}_{p_1h_1p_2h_2J_pJ_h} &=& \hspace{-7mm} \sum_{m_{p_1}m_{p_2}m_{h_1}m_{h_2}M_pM_h} 
    \hspace{-9mm} 
    \langle j_{p_1}m_{p_1}j_{p_2}m_{p_2} | J_pM_p\rangle 
    \nonumber \\ & &  \times 
    \langle j_{h_1}m_{h_1}j_{h_2}m_{h_2} | J_hM_h\rangle 
    \nonumber \\ & &  \times 
    (-1)^{J_j-M_h}  
    \langle J_pM_pJ_h-M_h | JM\rangle 
    \nonumber \\ & &  \times 
    (1+\delta_{p_1p_2})^{-1/2}(1+\delta_{h_1h_2})^{-1/2} 
    \nonumber \\ & &  \times 
    a_{p_1,m_{p_1}}^{\dagger} 
    a_{p_2,m_{p_2}}^{\dagger} 
    a_{h_1,m_{h_1}} 
    a_{h_2,m_{h_2}} 
\end{eqnarray} 
Henceforth the indices $JM$ will be omitted, but implied throughout. 
 
The $2p2h$ state (and the corresponding creation operator) is characterized, besides $JM$, by $J_p$ and $J_h$, the angular momenta to which the two 
particle states and the two hole states, respectively, are coupled. 
The same holds for the amplitudes $\mathcal{X}$, $\mathcal{Y}$.  
Moreover, an ordering of the single-particle states is introduced and only the operators with $p_1\leq p_2$, $h_1\leq h_2$ 
are included in the expansion (\ref{E:Qop}) to avoid multiple counting of configurations. 
For example, the present convention is that $\alpha < \beta$ if $t_{\alpha}<t_{\beta}$, or, 
for states of the same isospin, if $(\ell j)_{\alpha} < (\ell j)_{\beta}$, or, when all other quantum numbers are the same, if $n_{\alpha}<n_{\beta}$. 
  
The SRPA ground state, which formally is the vacuum of the annihilation operators $Q_{\lambda}$, 
is approximated with the Hartree-Fock (HF) ground state. 
The latter is the Slater determinant that minimizes 
the expectation value of the given Hamiltonian, $H=H_1+H_2$. 
The forward ($X$, $\mathcal{X}$) 
and backward ($Y$, $\mathcal{Y}$) 
amplitudes are the solutions of the SRPA equations  
in $ph\oplus 2p2h$ space 
\begin {equation} 
\left( \begin{array}{cc|cc}  
A                & \mathcal{A}_{12} & B  & 0 \\ 
\mathcal{A}_{21} & \mathcal{A}_{22} & 0  & 0 \\ \hline  
-B^{\ast}        &  0               & -A^{\ast} & -\mathcal{A}^{\ast}_{12} \\ 
   0             &  0               & -\mathcal{A}_{21}^{\ast} & -\mathcal{A}^{\ast}_{22} \\ 
\end{array} 
\right) 
\left( 
\begin{array}{c} 
X^{\lambda} 
\\ 
\mathcal{X}^{\lambda}  
\\ 
\hline  
Y^{\nu} 
\\ 
\mathcal{Y}^{\lambda}  
\end{array} 
\right) 
= E_{\lambda}  
\left( 
\begin{array}{c} 
X^{\nu} 
\\ 
\mathcal{X}^{\lambda}  
\\ 
\hline  
Y^{\nu} 
\\ 
\mathcal{Y}^{\lambda}  
\end{array} 
\right) 
\label{Esrpa}  
. \end{equation} 
The vanishing blocks are due to the choice of ground state. 
$A$ and $B$  
are the usual $N_1\times N_1$ RPA matrices ($N_1$ the number of $ph$ configurations), 
whose angular momentum-coupled forms are given by 
\begin{eqnarray} 
[A]_{ph;p'h'} 
&=& (e_p-e_h)\delta_{pp'}\delta_{hh'} 
    \nonumber \\ 
&&+ \langle ph^{-1};J | H_2 | p'{h'}^{-1};J\rangle  
\nonumber \\ 
&=& (e_p-e_h)\delta_{pp'}\delta_{hh'} 
    \nonumber \\ 
&&+ \sum_{J_1}(-1)^{j_{h}+j_{p'}-J_1} (2J_1+1)  
    \nonumber \\ && \times  
    \left\{  \begin{array}{ccc}j_{p} & j_{h'} & J_1\\ j_{p'} & j_{h} & J \end{array} \right\} 
    \langle ph';J_1 | H_2 | h p';J_1\rangle  
\label{E:Amat} 
\end{eqnarray} 
\begin{eqnarray} 
[B]_{ph;p'h'} 
&=& \langle (ph^{-1};J) (p'{h'}^{-1};J) | H_2 | 0\rangle  
\nonumber \\ 
&=& \sum_{J_1}(-1)^{j_{h}+j_{p'}+J-J_1} (2J_1+1)  
    \nonumber \\ && \times  
    \left\{  \begin{array}{ccc}j_{p} & j_{p'} & J_1\\ j_{h} & j_{h'} & J \end{array} \right\} 
    (1+\delta_{pp'})^{1/2} 
    (1+\delta_{hh'})^{1/2} 
    \nonumber \\ & &\times  
    \langle pp';J_1 | H_2 | h h';J_1\rangle  
. 
\label{E:Bmat} 
\end{eqnarray} 
The $N_1\times N_2$ submatrix $\mathcal{A}_{12}$ 
($N_2$ the number of $2p2h$ configurations) 
describes the coupling between $ph$ and $2p2h$ states,  
\begin{eqnarray} 
\lefteqn{
[\mathcal{A}_{12}]_{ph;p_1p_2h_1h_2J_pJ_h} 
= } \nonumber \\ 
&& \langle ph^{-1};J | H_2 | (p_1p_2;J_p)(h_1h_2;J_h)^{-1};J \rangle  
\nonumber \\ 
&=&  [1-(-1)^{j_{h_1}+j_{h_2}-J_h} P(h_1,h_2)] 
    \delta_{h_1h}
    \nonumber \\ & & \times  
    (-1)^{j_{p}+j_{h_2}+J+J_h}(1+\delta_{h_1h_2})^{-1/2}  
    \hat{J}_p\hat{J}_h 
    \nonumber \\ & & \times  
    \left\{  \begin{array}{ccc}J_p & J & J_h\\ j_{h_1} & j_{h_2} & j_p \end{array} \right\} 
    \langle p_1p_2;J_p | H_2 | ph_2;J_p\rangle  
    \nonumber \\ & - & 
  [1-(-1)^{j_{p_1}+j_{p_2}-J_p} P(p_1,p_2)] 
    \delta_{p_1p} 
    \nonumber \\ & & \times  
    (-1)^{j_{p_1}+j_{p_2}+J+J_h}(1+\delta_{p_1p_2})^{-1/2}  
    \hat{J}_p\hat{J}_h 
    \nonumber \\ & & \times  
    \left\{  \begin{array}{ccc}J_h & J & J_p\\ j_{p_1} & j_{p_2} & j_h \end{array} \right\} 
    \langle hp_2;J_h | H_2 | h_1h_2;J_h\rangle  
, 
\label{E:A12mat} 
\end{eqnarray}  
while 
the $N_2\times N_2$ matrix 
$\mathcal{A}_{22}$ contains the $2p2h$ states and their interactions, 
\begin{eqnarray} 
\lefteqn{
[\mathcal{A}_{22}]_{p_1p_2h_1h_2J_pJ_h;p_1'p_2'h_1'h_2'J_p'J_h'} 
= } \nonumber \\ 
&&  \delta_{p_1p_1'}\delta_{h_1h_1'}\delta_{p_2p_2'}\delta_{h_2h_2'}
     \delta_{J_pJ_p'}\delta_{J_hJ_h'}(e_{p_1}+e_{p_2}-e_{h_1}-e_{h_2}) 
    \nonumber \\ &+& 
    \langle 
    (p_1p_2;J_p)(h_1h_2;J_h)^{-1};J   
    | H_2 | 
    (p_1'p_2';J_p')({h_1'}{h_2'};J_h')^{-1};J   
    \rangle  
    \nonumber \\ 
&=&  \delta_{p_1p_1'}\delta_{h_1h_1'}\delta_{p_2p_2'}\delta_{h_2h_2'}
     (e_{p_1}+e_{p_2}-e_{h_1}-e_{h_2}) 
    \nonumber \\ &+& 
    \delta_{p_1p_1'}\delta_{p_2p_2'}   
    \delta_{J_pJ_p'}\delta_{J_hJ_h'} [1+(-1)^{J_p}\delta_{p_1p_2} ] (1+\delta_{p_1p_2})^{-1}  
    \nonumber \\ & & \times  
    \langle h_1'h_2';J_h | H_2 | h_1h_2;J_h\rangle  
    \nonumber \\ &+& 
    \delta_{h_1h_1'}\delta_{h_2h_2'}  
    \delta_{J_pJ_p'}\delta_{J_hJ_h'} [1+(-1)^{J_h}\delta_{h_1h_2} ] (1+\delta_{h_1h_2})^{-1}  
    \nonumber \\ & & \times  
    \langle p_1p_2;J_p | H_2 | p_1'p_2';J_p\rangle  
    \nonumber \\ &+& 
    [1-(-1)^{j_{p_1}+j_{p_2}-J_p} P(p_1,p_2)] (1+\delta_{p_1p_2})^{-1/2}  
    \nonumber \\ && \times  
    [1-(-1)^{j_{h_1}+j_{h_2}-J_h} P(h_1,h_2)] (1+\delta_{h_1h_2})^{-1/2}   
    \nonumber \\ && \times  
    [1-(-1)^{j_{p_1'}+j_{p_2'}-J_p} P(p_1',p_2')] (1+\delta_{p_1'p_2'})^{-1/2}  
    \nonumber \\ && \times  
    [1-(-1)^{j_{h_1'}+j_{h_2'}-J_h} P(h_1',h_2')] (1+\delta_{h_1'h_2'})^{-1/2}   
    \nonumber \\ && \times  
    \delta_{h_2h_2'}\delta_{p_2p_2'} (-1)^{1+j_{p_1}+j_{p_2}+j_{h_1}+j_{h_2}}   
    \hat{J}_p \hat{J}_p' \hat{J}_h \hat{J}_h'  
    \nonumber \\ && \times  
    \sum_{L}(-1)^{J_h-J_{h'}+J-L} (2L+1) 
    \left\{  \begin{array}{ccc}J_p & J_p' & L\\ J_h' & J_h & J \end{array} \right\} 
    \nonumber \\ && \times  
    \left\{  \begin{array}{ccc}J_p & J_p' & L\\ j_{p_1'} & j_{p_1} & j_{p_2} \end{array} \right\} 
    \left\{  \begin{array}{ccc}J_h & J_h' & L\\ j_{h_1'} & j_{h_1} & j_{h_2} \end{array} \right\} 
    \nonumber \\ && \times  
    \sum_{J_1}(-1)^{j_{h_1}+j_{p_1'}-J_1} (2J_1+1)  
    \nonumber \\ && \times  
    \left\{  \begin{array}{ccc}j_{p_1} & j_{h_1'} & J_1\\ j_{h_1} & j_{p_1'} & L \end{array} \right\} 
    \langle p_1h_1';J_1 | H_2 | p_1'h_1;J_1\rangle  
. 
\label{E:A22mat} 
\end{eqnarray}  
In the above, $e_{\alpha}$ are the HF single-particle energies 
and the operator $P(\alpha ,\beta )$ exchanges the indices $\alpha$ and $\beta$. 
The two terms in $\mathcal{A}_{12}$ describe the free propagation of the hole $h$ 
while the particle $p$ interacts with an intermediate $ph$ state
(terms with $h_1=h$ or $h_2=h$, i.e., containing $\delta_{h_{1,2}h}$), 
and the free propagation of a particle 
while the hole $h$ interacts with an intermediate $ph$ state 
(terms with $\delta_{p_{1,2}p}$). 
Thus self-energy corrections via bubble diagrams are introduced either to a particle or a hole state, or a $ph$ excitation is exchanged between a particle and a hole state. 
The four terms in $\mathcal{A}_{22}$ describe: the free propagation of a $2p2h$ state; the free propagation of the two particles while the two holes interact; the free propagation of the two holes while the two particles interact; and the free propagation of a $ph$ pair while the other particle and hole interact. 
In the following we will omit the $J_p$, $J_h$ indices for simplicity. 
 
If we neglect the coupling amongst the $2p2h$ states, $\mathcal{A}_{22}$ becomes diagonal and its elements are determined by the 
unperturbed $2p2h$ energies (diagonal approximation), 
\begin{eqnarray} 
\lefteqn{[\mathcal{A}_{22}]_{p_1h_1p_2h_2,p_1'h_1'p_2'h_2'} = } 
\nonumber \\ 
&& 
 \delta_{p_1p_1'}\delta_{h_1h_1'}\delta_{p_2p_2'}\delta_{h_2h_2'}
(e_{p_1}+e_{p_2}-e_{h_1}-e_{h_2}) 
\label{Ea22} 
.  
\end{eqnarray} 
The validity of the diagonal approximation is examined in Sec.~\ref{S:Diag}. 

As long as we are interested only in the single-particle response, determined by the $ph$ amplitudes, $X^{\lambda}$ and $Y^{\lambda}$, we may eliminate the $\mathcal{X}^{\lambda}$, $\mathcal{Y}^{\lambda}$ amplitudes from Eq.~(\ref{Esrpa}) and reduce the SRPA problem to an energy-dependent eigenvalue problem of the dimension of the RPA matrix~\cite{daP1965},  
\begin{equation} 
\left( \begin{array}{cc}  
      A(E_{\lambda} )  & B \\ -B^{\ast} & -A^{\ast}(-E_{\lambda} )  
\end{array} \right) 
\left( \begin{array}{cc}  
      X^{\lambda} \\ Y^{\lambda} 
\end{array} \right) 
 = E_{\lambda}  
\left( \begin{array}{cc}  
      X^{\lambda} \\ Y^{\lambda} 
\end{array} \right) 
. 
\label{E:SRPAred1} 
\end{equation} 
In general, the expression for $A_{php'h'}(E)$ will involve the inverse of $[(E+i\eta_2)\mathcal{I}_{N_2} - \mathcal{A}_{22}]$, where $\mathcal{I}_{N_2}$ the $N_2\times N_2$ identity matrix. 
Within the diagonal approximation we have simply  
\begin{eqnarray} 
\lefteqn{
A_{php'h'}(E ) =  A_{php'h'} 
 } \nonumber \\ 
&& + \sum_{p_1p_2h_1h_2} 
      \frac{[\mathcal{A}_{12}]_{ph;p_1p_2h_1h_2}[\mathcal{A}^T_{12}]_{p_1p_2h_1h_2;p'h'}}{E - (e_{p_1} + e_{p_2} - e_{h_1}-e_{h_2}) + i\eta_2} .  
\label{E:SRPAred2} 
\end{eqnarray} 
A finite constant $\eta_2>0$ is used in applications to smoothen the poles of the function and to introduce a width to the $2p2h$ states. 

\subsection{Quantities of interest} 

\label{S:QOI} 

The quantities of interest are transition strength distributions, or strength functions, $R_F(E)$ 
of transition operators ${F}^{\dagger}$ 
\begin{eqnarray} 
R_{F}(E) &=&  \sum_{\lambda} |\langle \lambda | {F}^{\dagger} |0 \rangle |^2 \delta (E-E_{\lambda})  \\ 
                     &\equiv& \sum_{\lambda} B_F(E_{\lambda}) \delta (E-E_{\lambda}) 
,
\label{E:Rdiscr}
\end{eqnarray} 
and their energy moments 
\begin{equation} 
m_k=\sum_{\lambda}E_{\lambda}^k B_F(E_{\lambda})  , 
\label{E:moments} 
\end{equation} 
determined, in general, by the amplitudes $X$, $Y$, $\mathcal{X}$, and $\mathcal{Y}$ through 
\begin{eqnarray} 
\lefteqn{\langle \lambda | F^{\dagger} |0 \rangle 
= 
     \sum_{ph} [ f_{ph}{X_{ph}^{\lambda}}^{\ast} - \tilde{f}_{ph}{Y_{ph}^{\lambda}}^{\ast}] 
}
\nonumber \\ 
&&+ 
  \!\!\! \sum_{p_1p_1h_1h_2} \!\!\!  
   [ f_{p_1p_2h_1h_2}{\mathcal{X}_{p_1p_2h_1h_2}^{\lambda}}^{\ast} - \tilde{f}_{p_1p_2h_1h_2}{\mathcal{Y}_{p_1p_2h_1h_2}^{\lambda}}^{\ast}] 
,
\label{E:Fl0} 
\end{eqnarray} 
where the coefficients $f$, $\tilde{f}$ depend on the operator and ground state. 
Centroid energies can be defined as 
\begin{equation} 
\bar{E} \equiv m_1/m_0, 
\end{equation} 
where the sums $m_k$ may be evaluated over the whole spectrum 
(unrestricted summations in Eq.~(\ref{E:moments})), or only in the energy region of a resonance. 
The width of a distribution or a resonance can be expressed as 
\begin{equation} 
\Delta \equiv \sqrt{\frac{m_2}{m_0} - \bar{E}^2}  
\label{E:width} 
.  
\end{equation}  

Smoothened strength functions can be produced, for analysis purposes, by folding the discrete strength functions with a Lorentzian of width $\Gamma$, which yields 
\begin{eqnarray} 
R_{F}(E) &=&  \frac{1}{2\pi }\sum_{\lambda} B_F(E_{\lambda})\frac{\Gamma}{(E-E_{\lambda})^2+\Gamma^2/4} 
.
\label{E:Lor} 
\end{eqnarray} 
In the limit $\Gamma\rightarrow 0$, Eq.~(\ref{E:Rdiscr}) is recovered.  

We will consider IS and IV transitions of definite spin and parity $J^{\pi}$,  
described by standard single-particle transition operators~\cite{PPH2006}. 
Then, for the HF ground state, we have 
\begin{equation} 
B_F(E_{\lambda}) 
 = \frac{1}{{2J+1}} | \sum_{ph} [{X_{ph}^{\lambda}}^{\ast}+(-1)^J {Y_{ph}^{\lambda}}^{\ast}] 
   \langle p || F || h \rangle |^2 
. 
\end{equation} 
The two-body $\mathcal{X}$, $\mathcal{Y}$ amplitudes do not contribute. 
The energy moments $m_0$ and $m_1$ will be the same in SRPA as in RPA~\cite{AdL1988}. 
A $(2J+1)$ multiplicity will be included in our final results.  

Two-body operators can also be considered. 
Then all amplitudes contribute to the transition matrix element. 
For example,   
let us consider the double dipole resonance, excited by the two-body operator $F_{\mathrm{DDR};J}=[F_{\mathrm{IVD}}\otimes F_{\mathrm{IVD}}]_{J^+}$, where $F_{\mathrm{IVD}}$ the usual (single-particle) isovector $1^-$ operator and $J=0,2$. 
We then have 
\begin{eqnarray} 
\lefteqn{\langle \lambda | F_{\mathrm{DDR};J}^{\dagger} |0 \rangle 
 = \sum_{ph} [{X_{ph}^{\lambda}}^{\ast}+(-1)^J {Y_{ph}^{\lambda}}^{\ast}] 
   f_{ph}^{\mathrm{DDR};J} }
\nonumber \\ 
&& + \!\!\! \sum_{p_1p_2h_1h_2} \!\!\! 
   [{X_{p_1p_2h_1h_2}^{\lambda \, \, \ast}}+{Y_{p_1p_2h_1h_2}^{\lambda \, \, \ast}}] 
   f_{p_1p_2h_1h_2}^{\mathrm{DDR};J} 
, 
\end{eqnarray} 
where~\cite{Nis0000}  
\begin{eqnarray} 
f_{ph}^{\mathrm{DDR};J}  
&=& (-1)^{j_p+j_h+J} \sum_{p'}  
  \left\{   \begin{array}{ccc} 1 & 1 & J \\ j_h & j_p & j_{p'} \end{array}   \right\} 
\nonumber \\ 
&& \times \langle p || F_{\mathrm{IVD}} || p' \rangle  
   \langle p'|| F_{\mathrm{IVD}} || h  \rangle 
\nonumber \\  
&+& (-1)^{j_p+j_h+1} \sum_{h'}  
  \left\{   \begin{array}{ccc} 1 & 1 & J \\ j_p & j_h & j_{h'} \end{array}   \right\} 
\nonumber \\ 
&& \times \langle p || F_{\mathrm{IVD}} || h' \rangle  
   \langle h'|| F_{\mathrm{IVD}} || h  \rangle 
\end{eqnarray} 
and 
\begin{eqnarray} 
\lefteqn{f_{p_1p_2h_1h_2}^{\mathrm{DDR};J} = 
      2\sqrt{\frac{(2J_p+1)(2J_h+1)}{(1+\delta_{p_1p_2})(1+\delta_{h_1h_2})}}} 
\nonumber \\ 
&& \times \left[ 
   \langle p_1 || F_{\mathrm{IVD}} || h_1  \rangle 
   \langle p_2 || F_{\mathrm{IVD}} || h_2  \rangle 
   \left\{ 
     \begin{array}{ccc} 
     j_{p_1}   &   j_{h_1}   &    1   \\ 
     j_{p_2}   &   j_{h_2}   &    1   \\ 
     J_p       &   J_h       &    J    
     \end{array} 
   \right\}
   \right. 
\nonumber \\ 
&& - (-1)^{j_{h_1}+j_{h_2}-J_h}  
\nonumber \\ 
&& 
   \mbox{~}\,\,\,\,  
   \times 
   \langle p_1 || F_{\mathrm{IVD}} || h_2  \rangle 
   \langle p_2 || F_{\mathrm{IVD}} || h_1  \rangle 
\nonumber \\ 
&& 
   \mbox{~}\,\,\,\,  
   \times 
   \left. 
   \left\{ 
     \begin{array}{ccc} 
     j_{p_1}   &   j_{h_2}   &    1   \\ 
     j_{p_2}   &   j_{h_1}   &    1   \\ 
     J_p       &   J_h       &    J    
     \end{array} 
   \right\}
   \right]  
.
\end{eqnarray} 

Finally, we may define the strength distribution of a $|ph^{-1}\rangle$ configuration, coupled to a given angular momentum state, 
via the quantity 
\begin{eqnarray} 
S_{ph}(E) 
&=& \sum_{\lambda >0} (|X_{ph}^{\lambda}|^2 - |Y_{ph}^{\lambda}|^2 ) \delta (E-E_{\lambda}) 
\nonumber \\ 
&\equiv & \sum_{\lambda >0} s_{ph}(E_{\lambda})  \delta (E-E_{\lambda}) 
,
\label{Esph}
\end{eqnarray}  
where the summation is over all eigenstates with $E_{\lambda}>0$. 
(In the unperturbed case the centroid of $S_{ph}(E)$ is trivially identical to the HF $ph$ energy $e_p-e_h$ and its width is zero.) 
Similarly, the strength distribution of a $2p2h$ state can be defined using the $\mathcal{X}$ and $\mathcal{Y}$ amplitudes. 
Energy moments and centroids, as well as smoothened distributions, can be defined as usual. 

We note that the total $m_0$ and $m_1$ (and centroid) of $S_{ph}$ will be the same in RPA and SRPA, since $S_{ph}$ is the sum of the strength functions related to the operator ${O^{JM}_{ph}}^{\dagger}$ and its adjoint. 
$m_0$, in particular, should always amount to one. 

\subsection{Related approximations and ground-state correlations} 

\label{S:rsrpa} 

By setting the coupling matrices $\mathcal{A}_{12}$ and $\mathcal{A}_{22}$ and the $2p2h$ amplitudes $\mathcal{X}$, $\mathcal{Y}$ equal to zero  
in Eq.~(\ref{Esrpa}), we recover the usual RPA problem. If, in addition, we neglect the $ph$ residual interaction (i.e., $B_{ph,p'h'}=0$ and 
$A_{ph,p'h'}=(e_p-e_h)\delta_{pp'}\delta_{hh'}$), we obtain a trivial, unperturbed problem, 
where the eigenstates $|\lambda\rangle$ are the $ph$ configurations at the HF level and the $Y$ amplitudes vanish. 

By setting only $B=0$ 
in Eq.~(\ref{Esrpa}), we obtain a second-order Tamm-Dancoff approximation (STDA), which amounts to solving the eigenvalue problem of the $A$ block of the SRPA matrix. The backward amplitudes $Y$ and $\mathcal{Y}$ vanish and ground-state correlations implicitly taken care of by those are ignored. Setting also 
the coupling matrices $\mathcal{A}_{12}$ and $\mathcal{A}_{22}$ equal to zero, one gets the usual, first-order Tamm-Dancoff approximation (TDA).   

In a manner analogous to TDA, STDA is equivalent to a diagonalization of the Hamiltonian in the 
$ph\oplus 2p2h$ space. 
It should be noted, though, that for $J^{\pi} = 0^+$ the HF ground state does not decouple from the STDA space (unlike TDA). A diagonalization in the model space that includes in addition the HF state would produce a new ground state of lower energy. The $0^+$ excitation spectrum would also be affected.  

In Sec.~\ref{S:cons} we will investigate the possible influence of ignored ground-state correlations (GSC) on our results. A rigorous way to do that would be the use of an extended SRPA method built on a correlated ground state as self-consistently as possible (or other extended methods such as a self-consistent Green's function method~\cite{BaD2003}). Since, however, that is a demanding project going beyond the scope of the present work, we resort instead to simpler approaches. 
One way to assess the role of GSC is to ignore them completely. This is accomplished within the (S)TDA. As a second approach, we have devised a simple and rather heuristic renormalized version of SRPA (RSRPA), which takes into account to some extent the depletion of the Fermi sea. 

We start with a renormalized RPA approach (RRPA). Following the simplified RRPA method of Refs.~\cite{CPS1996,VKC2000} --- see also Ref.~\cite{PRP2007} --- 
we assume partially occupied single-particle states and renormalize the residual couplings and the transition matrix elements accordingly. In particular, when calculating the matrix elements of $A$, the $H_2$ terms in Eq.~(\ref{E:Amat}) are multiplied by a factor 
\begin{equation}
     d_{ph,p'h'}^{[1]} = D_{ph}^{1/2}D_{p'h'}^{1/2} \, ,  
\label{E:Dfact1} 
\end{equation} 
where 
\begin{equation}
     D_{ph} \equiv n_h-n_p 
\label{E:Dph} 
\end{equation} 
and $n_{\alpha}$ is the occupation probability of the orbital $\alpha$. 
The matrix elements of $B$, Eq.~(\ref{E:Bmat}),  are multiplied by the same factor. 
Finally, the single-particle transition matrix elements $f_{ph}$, $\tilde{f}_{ph}$, are renormalized by a factor $D_{ph}^{1/2}$. 

In RSRPA we have to renormalize the matrix elements of $A_{12}$ and $A_{22}$ as well. 
The former, Eq.~(\ref{E:A12mat}), will be multiplied by the expression (\ref{E:Dfact1}), taking account of the occupation probabilities of the $p$, $h$ states, 
as well as by a factor 
\begin{equation}
     d_{p_1h_1p_2h_2}^{[2]} = \frac{1}{2}[ d_{p_1h_1,p_2h_2} + d_{p_1h_2,p_2h_1}]  
, 
\label{E:Dfact2} 
\end{equation} 
to take into account the occupation probabilities of the $p_{1,2}$, $h_{1,2}$ states. 
Similarly, the $H_2$ terms in the $A_{22}$ matrix elements, Eq.~(\ref{E:A22mat}), will be multiplied by 
$ d_{p_1h_1p_2h_2}^{[2]} d_{p_1'h_1'p_2'h_2'}^{[2]} $.   

In practice, we will not solve these equations iteratively. The single-particle energies and eigenstates needed to evaluate the above matrix elements will be the HF ones, whereas the occupation probabilities are calculated using the shell model. Obviously, this method is neither rigorous nor consistent, but it should help us get an idea regarding the sensitivity of our results to GSC. 

\subsection{Consistency and stability} 

\label{S:cons1} 
 
It is well known, that in self-consistent RPA (meaning that the same Hamiltonian is used to calculate the HF ground state and the residual interaction) the spurious state related to the CM momentum operator $\hat{\vec{P}}$ (single spurious state) will appear at zero energy and be exactly separated from the physical spectrum, provided that all $ph$ and $hp$ configurations available in the (sufficiently large) single-particle space are taken into account. The same does not hold, however, when extensions of RPA are considered, whether that means considering a correlated ground state or higher-order configurations. As was formally shown in Ref.~\cite{ToS2004}, once $2p2h$ configurations are included in the model space, a necessary condition for the single spurious state to appear at zero energy is that   
all single-particle amplitudes be taken into account: This means not only $ph$ and $hp$ amplitudes, but also $pp$ and $hh$. 
It was in fact shown that $pp$ and $hh$ amplitudes affect the spurious state even with a HF ground state, because at energies equal to $e_p-e_{p'}$ or $e_h-e_{h'}$ they do not vanish. 
Obviously our approach does not include them. This problem will be examined quantitatively in Sec.~\ref{S:cons}. 

A more severe problem in HF-based SRPA is the onset of instabilities. 
Contrary to RPA, the ``self-consistent" use of the HF ground state 
does not guarantee that the SRPA stability 
matrix will be positive-definite: 
Thouless's theorem was proven specifically for $ph$ excitations~\cite{Tho1960,Tho1961}. 
Thus RPA is related to the stability conditions for the HF solution, but SRPA seems 
related to 
an extended variational problem~\cite{daP1965,DNS1990}. 
In practice, we will find that low-lying states 
appear at imaginary or negative energies.%
\footnote{Here we speak of a ``negative-energy" excitation when the eigenstate with $E_{\lambda}<0$ is normalized to +1, i.e., the norm of the positive-energy counterpart is negative.}  
It is not difficult to demostrate how negative eigenvalues can occur, already at the STDA level ($B=0$), if we consider Eq.~(\ref{E:SRPAred2}): 
For energies $E$ below typical $2p2h$ energies, strong (or many) $\mathcal{A}_{12}$ elements, regardless of sign, can cause even the diagonal elements of $A(E)$ to become negative and destroy the positive-definitness of the matrix.  
Additional ground-state correlations could cure this problem, through a renormalization of the $\mathcal{A}_{12}$ couplings, as illustrated above, or by filling the vanishing $B-$submatrices of Eq.~(\ref{Esrpa}) with finite elements (see also Ref.~\cite{BaD2003} for a related discussion within an extended dressed RPA). 
 
The question whether SRPA gives meaningful results for giant resonances despite the problematic solutions at low energies is also tackled in Sec.~\ref{S:cons}.

\section{Solving the SRPA problem} 

\label{S:solve} 

In practice we proceed as follows. We first choose a single-particle space, consisting of harmonic-oscillator eigenstates. The larger the space, the better the convergence of our results to their final values. 
All angular-momentum coupled two-body matrix elements 
of the given interaction within the harmonic-oscillator basis 
are calculated in advance as described in Ref.~\cite{RHP2005} and stored. 
We solve the HF equations within the given space and then we use all $ph$ and $2p2h$ configurations, which are available within the space and can couple to the desired angular momentum and parity, to construct the SRPA matrix. 
The single-particle space is characterized by either 
\begin{itemize} 
\item{ 
$\epsilon_{\max}=(2n+\ell )_{\max}$, the number of energy quanta in the highest oscillator shell considered. This means that all single-particle states within the lowest $\epsilon_{\max}+1$ shells are used. An additional cutoff $\ell_{\max}$ in $\ell$ may also be imposed. 
} 
\end{itemize} 
or  
\begin{itemize} 
\item{ 
$n_{\max}$ and $\ell_{\max}$. A cutoff $\ell_{\max}$ in $\ell$ is considered and each possible $\ell j$ state is expanded in the lowest $n_{\max}+1$ harmonic-oscillator $\ell$ states. 
} 
\end{itemize} 
Since angular momentum is preserved, 
for an RPA calculation of the $J^{\pi}$ response, one has to include $j$'s up to the maximal occupied one plus $J$, so that all $j_p$'s with $|j_h-j_p|\leq J\leq j_p+j_h$ are included in the space, and may omit higher $j_p$'s. 
In SRPA there is no $j$-cutoff provided by angular momentum conservation. In principle, particle states with infinitely large $j$'s can contribute to allowed $2p2h$ configurations and such configurations can couple with $ph$ ones through finite matrix elements of $\mathcal{A}_{12}$.
In particular, infinitely large $j_{p_1}$ and $j_{p_2}$ can couple to any given $J_p$, thus resulting in finite values of 
    $\langle p_1p_2;J_p | H_2 | ph_2;J_p\rangle$  
--- see r.h.s. of Eq.~(\ref{E:A12mat}).  
For restricted $n_{\max}$, however, upper bounds are provided by the transformation (Moshinsky) brackets needed to evaluate the above matrix element in the HO basis~\cite{RPP2006}. 
Natural cutoffs on $n_{\max}$ or $\epsilon_{\max}$ should be provided by the properties of the interaction. 
Perhaps such considerations can be used, along with the convergence behavior of the SRPA solutions with respect to the single-particle basis, for the optimization of the latter, but so far this has not been done. 

The next task is to solve the SRPA eigenvalue problem. 
First we note that all the submatrices comprising the SRPA matrix, Eq.~(\ref{Esrpa}), are real in the cases studied here. The submatrices $A$, $B$, and $\mathcal{A}_{22}$ are also symmetric. The same holds for the $N\times N$ blocks indicated in Eq.~(\ref{Esrpa}) (separated by lines). 
The dimension $2N$ of the SRPA matrix is given by twice the number of $ph$ and of $2p2h$ configurations available in the model space ($N=N_1+N_2$). 
The latter, $N_2$, can be rather large: For the purposes of the present work we encountered problems with $N$ up to $10^6$. In the following we present technical information on how we deal with the SRPA eigenvalue problem in practice. We shall distinguish three cases: 
\begin{itemize}
\item{ 
All eigenvalues and eigenvectors are calculated. This is feasible for relatively small problems, e.g., $N\sim 10^{4}$. 
} 
\item{ 
For large problems the complete solution becomes impossible. Then only a small portion of consecutive eigenvalues and the respective eigenvectors are calculated. In most cases, all excitations of interest lie at the lower end of the spectrum, so it suffices to evaluate only the lowest (a few tens or hundreds) positive eigenvalues. 
}  
\item{ 
No eigenvalue problem is solved. The response function is evaluated directly. The strategy is well known and very practical when one is interested in the single-particle response, and only in the final result, namely a smoothened strength function. 
} 
\end{itemize} 

\subsubsection{Small problems: all eigenvalues} 

Standard library routines can be used to solve small enough eigenvalue problems. The solution can be sped up considerably if the SRPA problem is reduced to half the dimension, $N\times N$. A method relying on a Cholesky decomposition of the matrix $A+B$ or $A-B$ is the most common way to do it, with the additional bonus that it produces a {\em symmetric} $N\times N$ eigenvalue problem~\cite{Chi1970}. However, the method will not work if neither $A+B$ nor $A-B$ is positive-definite, a problem that can occur when there are instabilities, or just spurious states at imaginary energies. We employ a modified method instead, relying on a generalized Cholesky decomposition~\cite{Pap2007}, which is equivalent to the original one when the matrix under decomposition is positive definite (thus still producing a symmetric eigenvalue problem) and still works when it is not (producing a non-symmetric eigenvalue problem of half the dimension), with minimal additional computational effort. 
For details, see Ref.~\cite{Pap2007}. 

\subsubsection{Large problems} 
When the dimension of the matrix is very large, the first problem that occurs is the storage of the matrix elements. Fortunately, most of the SRPA matrix matrix elements vanish, especially when the diagonal approximation is used, Eq.~(\ref{Ea22}), so it becomes possible to store all the finite ones in memory.  
Obviously, one needs to store only one $N\times N$ matrix (e.g., $A$ or $A\pm B$) and one small, $N_1\times N_1$ matrix ($B$). Moreover, only the upper (or lower) triangles need be stored, since the matrices are symmetric. 

The dimension and degree of sparseness of the SRPA matrix depends on the nucleus, type of response, and model space. As an example, let us mention that in the case of the $2^{+}$ response of $^{40}$Ca with the $V_{\mathrm{UCOM}}$, and for spaces large enough for reasonable convergence, there are about a million configurations. $0.5-1$GB of storage are needed in the diagonal approximation, but $100$~GB could be required for the full problem ($10^{10}$ elements in double precision). 

Again the dimension of the SRPA problem can be reduced by half, to speed 
up the numerical solution. 
In order to save matrix operations we chose not to perform a decomposition, but resort to a more straightforward reduction method~\cite{Pap2007}:  
We solve the non-symmetric $N\times N$ problem 
\begin{equation}  
      (A-B)(A+B)R^{\lambda} = E^2_{\lambda } R^{\lambda} \, ,  
\label{E:AB} 
\end{equation}  
where the eigenvectors $R^{\lambda} = E_{\lambda  }^{-1/2}(X^{\lambda} + Y^{\lambda})$ obey the 
normalization condition $(R^{\lambda})^T(A+B)R^{\mu}=\pm\delta_{\lambda\mu}$ (real and positive $E_{\lambda  }$). 
The properly normalized $X$ and $Y$ arrays ($|X^{\lambda}|^2-|Y^{\lambda}|^2=\pm 1$) are then given by 
\begin{eqnarray} 
X^{\lambda} &=& \frac{1}{2} [ \sqrt{E_{\lambda  }} \mathcal{I}_N + \frac{1}{\sqrt{E_{\lambda  }}} (A+B) ] R^{\lambda} 
\\ 
Y^{\lambda} &=& \frac{1}{2} [ \sqrt{E_{\lambda  }} \mathcal{I}_N - \frac{1}{\sqrt{E_{\lambda  }}} (A+B) ] R^{\lambda} 
, 
\end{eqnarray}  
where $\mathcal{I}_N$ is the $N\times N$ identity matrix. 

Finally, an Arnoldi iteration procedure from the ARPACK package~\cite{LSY1998} is employed to solve the problem (\ref{E:AB}) only for the $k$ lowest positive eigenvalues $E_{\lambda}^2$, where $k< < N$. 
In principle, it is possible to solve for the first $k$ eigenvalues lying above a given energy value $E_{\mathrm{offset}}^2$, not necessarily equal to zero. 
In practice, however, such a strategy can prove problematic: Already at moderate energies the density of eigenstates can be so large (see Sec.~\ref{S:res}), that the Arnoldi iteration will likely fail to converge. 


\subsubsection{Response function formalism} 

As long as one-body transition operators are considered, we may just solve the reduced SRPA problem, 
Eq.~(\ref{E:SRPAred1}). 
In general, the reduction procedure involves the inversion of a large matrix (of the dimension of the $2p2h$ space), but when $\mathcal{A}_{22}$ is diagonal, that is reduced to a trivial number inversion, see Eq.~(\ref{E:SRPAred2}). 
There are ways to solve such an energy-dependent eigenvalue problem~\cite{BAD1990,All1993}.  
An efficient alternative is to employ the response-function formalism. 
Then, instead of explicitly solving the eigenvalue problem, one can obtain directly the strength function of interest \cite{Wam1988,All1993}. 
First, the $ph$ Green's function, a $N_1\times N_1$ matrix, is evaluated, 
\begin{equation} 
 G(E ) = - \left( 
   \begin{array}{cc} 
   A(E ) - (E + i\eta_1)\mathcal{I}_{N_1} & B \\ 
                   B^{\ast}                  & A^{\ast}(-E )+(E -i\eta_1 )\mathcal{I}_{N_1}  
   \end{array} 
   \right)^{-1} 
\end{equation} 
($\eta_1\to 0^+$) by numerical matrix inversion. The response function for a given single-particle field is given by 
\begin{equation} 
 \mathcal{R}(E ) = (\mathcal{F}^T \, , \, \tilde{\mathcal{F}}^T ) 
                   G(E ) 
                   \left( 
                    \begin{array}{c} 
                          \mathcal{F} \\ \tilde{\mathcal{F}} 
                    \end{array} 
                   \right) 
\, ,  
\end{equation} 
where the elements of the $N_1-$dimensional (assumed real) array $\mathcal{F}$ ($\tilde{\mathcal{F}}$), in $ph$ space, are the matrix elements of the transition operator, $f_{ph}$ ($-\tilde{f}_{hp}$) --- see Eq.~(\ref{E:Fl0}). 
Finally,  the strength function of interest is given by 
\begin{equation} 
  R(E ) = -\frac{\Im}{\pi} \mathcal{R}(E ) \, .  
\end{equation} 

In practice, the Green's function, response function, and finally the strength function are evaluated over the energy range of interest, which is represented by mesh points $E_i$. The constants $\eta_1$, which provide the $ph$ and $2p2h$ states, respectively, with a finite width should, in principle, be small enough for the structure of the strength function to be resolved as desired. The mesh size $\delta E=E_{i+1}-E_i$ should be smaller than both of them (a factor 3-4 suffices). The choice $\eta_1=\eta_2$ produces a smoothened strength function that is practically the same as the discretized strength function (obtained by explicit diagonalization of the SRPA matrix) folded with a Lorentzian of width $\Gamma =2\eta_{1,2}$, cf. Eq.~(\ref{E:Lor}). 

\section{Results} 

\label{S:res} 

In the following, we will discuss the features of the SRPA solutions with the help of illustrative examples. 
We note that we will use the acronym SRPA0 when referring specifically to the diagonal approximation, Eq.~(\ref{Ea22}), and SRPA(0) when referring explicitly to both solutions, with and without the diagonal approximation. 
The intrinsic nuclear Hamiltonian is employed. 
In all cases we set the oscillator length parameter of the single-particle basis 
equal to $b=1.7$~fm (for $V_{\mathrm{UCOM}}$) or $b=1.8$~fm (for $V_{\mathrm{BB}}$).  

It will prove instructive to consider, among others,  
a relatively small SRPA problem, for which we can calculate all eigenstates --- with and without the diagonal approximation. 
The purpose of such an exercise is 
not to perform a realistic calculation, 
but to illustrate and discuss selected features of the method: 
in particular, how the large amount of $2p2h$ configurations influences the response function and the distribution 
of the SRPA eigenstates. 

\begin{figure*}[!]\centering
\includegraphics[angle=-90,width=5.8cm]{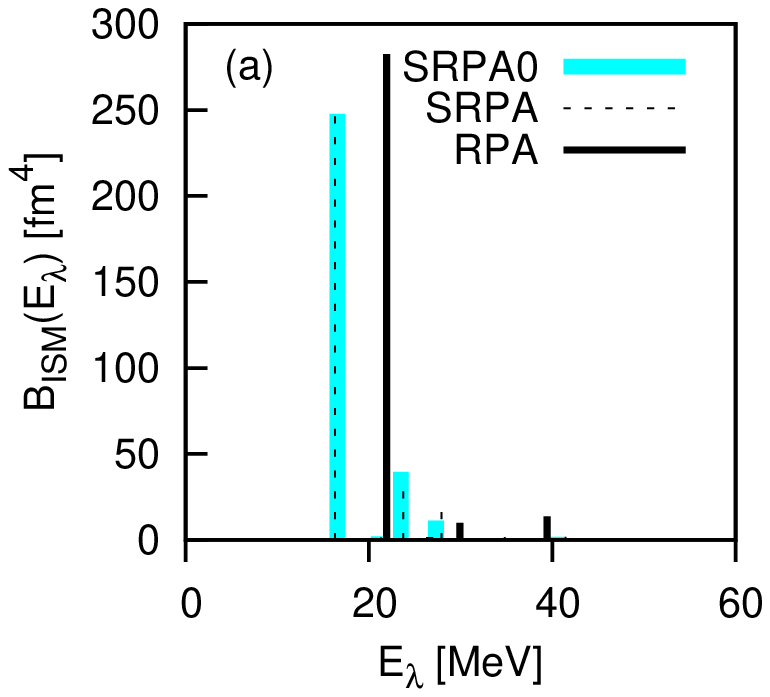} 
\includegraphics[angle=-90,width=11.6cm]{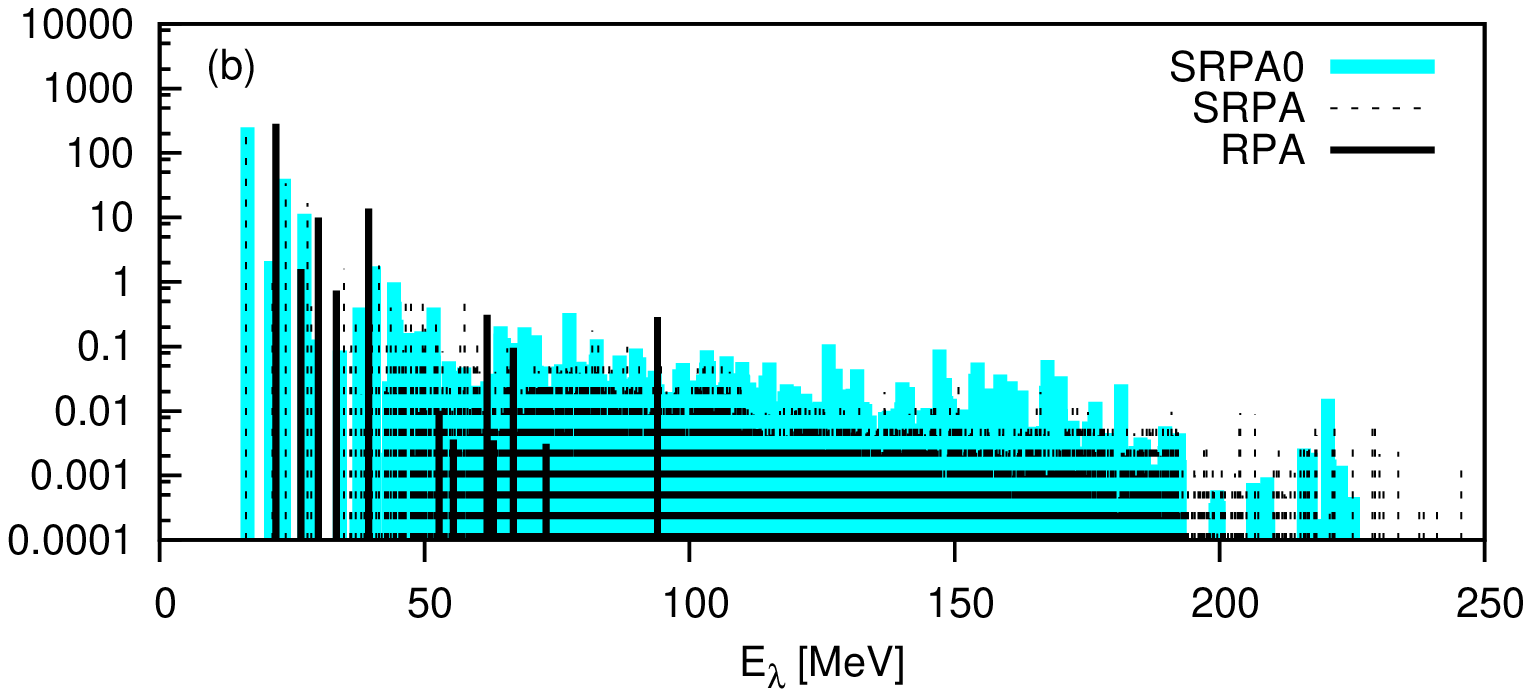} 
\caption{%
Isoscalar monopole response of $^{16}$O, calculated within a single-particle basis of 7 oscillator shells: 
Strength function calculated within SRPA and SRPA0 (diagonal approximation, Eq.~(\ref{Ea22})), as well as RPA, in (a) linear and (b) logarithmic scale. 
\label{F:toy}}
\end{figure*}

As such a ``toy model" we choose the monopole ($0^+$) response of $^{16}$O in a rather small single-particle space, consisting of seven oscillator shells, and using the $V_{\mathrm{UCOM}}$ interaction.  
In total, there are 4148 positive-energy eigenstates to evaluate 
($N_1=14$, $N_2=4134$).  
The single-particle, isoscalar monopole strength distribution is shown in Fig.~\ref{F:toy} in linear (Fig.~\ref{F:toy}a) and logarithmic (Fig.~\ref{F:toy}b) scale. 

Further examples will be introduced in the following as needed. 


\subsection{Diagonal approximation} 
\label{S:Diag} 

\begin{figure}[!]\centering
\includegraphics[angle=-90,width=8cm]{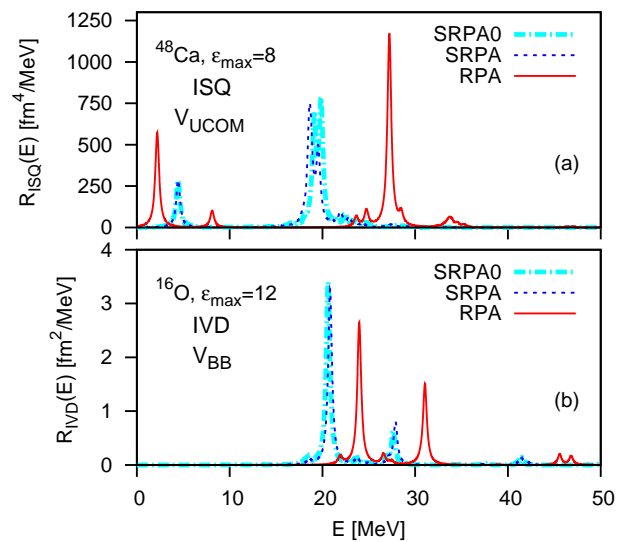} 
\caption{%
Quality of the diagonal approximation. {(a)}: IS quadrupole response of $^{48}$Ca in a single-particle basis with $\epsilon_{\max}=8$ and using $V_{\mathrm{UCOM}}$. {(b)}: IV dipole response of $^{16}$O in a single-particle basis with $\epsilon_{\max}=12$ and $\ell_{\max}=8$ and using $V_{\mathrm{BB}}$. 
Results are shown obtained with RPA, solving the full SRPA problem and using the diagonal approximation (SRPA0). 
(In all cases, $\Gamma =0.5$~MeV.)  
\label{F:full}}
\end{figure}

When looking at Fig.~\ref{F:toy}a, we observe that the results of the diagonal approximation, SRPA0, are very close to the exact SRPA results. 
It looks as though the effect of the $2p2h$ space on the single particle response is approximately the same, whether or not the $2p2h$ states are considered unperturbed. 
We have verified that the approximation is equally good in larger spaces and for different types of response --- always of single-particle operators. 
See, e.g., the IVD response of $^{16}$O calculated within a space of 13 oscillator shells, as shown in Fig.~1 of Ref.~\cite{PaR2009}. The approximation remains quite good for heavier nuclei, for example $^{48}$Ca, whose ISQ response is shown in Fig.~\ref{F:full}a, calculated within a space of 9 shells. 

We expect the diagonal approximation to be reliable for soft, perturbative interactions in general. 
The additional couplings within the $2p2h$ space, ignored in the diagonal approximation, constitute higher-order corrections to the excitation propagator with respect to the interaction, as can be demonstrated diagrammatically~\cite{Wam1988}.  
Results with the Brink-Boeker potential, $V_{\mathrm{BB}}$, corroborate this speculation.  
$V_{\mathrm{BB}}$ is even softer than the $V_{\mathrm{UCOM}}$. For example, it produces much smaller second-order corrections to the nuclear binding energies in the perturbation expansion beyond Hartree-Fock~\cite{Rot0000}. 
In Fig.~\ref{F:full}b we show the IVD response of $^{16}$O calculated using $V_{\mathrm{BB}}$. 
The diagonal approximation is very good in this case.  

To be on the safe side, it is always advisable to verify the quality of the approximation within some tractable space. 
In Ref.~\cite{GaC2009}, for example, it was found that it is quite bad in the case of metallic clusters when the bare 
Coulomb interaction is used. 

\begin{figure}[!]\centering
\includegraphics[angle=-90,width=8cm]{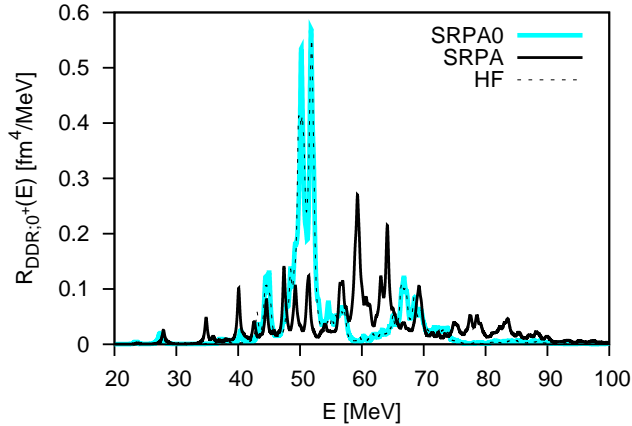} 
\caption{%
$0^+$ component of the double dipole resonance of $^{16}$O, calculated within a single-particle basis of 7 oscillator shells: 
Strength function ($\Gamma =0.5$~MeV) calculated within SRPA and SRPA0 (diagonal approximation, Eq.~(\ref{Ea22})), as well as using the unperturbed (HF) $ph$ and $2p2h$ states. 
\label{F:ddr}}
\end{figure}

From Fig.~\ref{F:toy}a we deduced that the SRPA and SRPA0 solutions yield almost the same results for the giant resonance. However, when we look at the strength distribution on a logarithmic scale, Fig.~\ref{F:toy}b, it becomes obvious that the two solutions give different results for the majority of eigenstates, which are mostly of $2p2h$ character. One implication is that the diagonal approximation cannot be relied upon when examining, e.g., double giant resonances and the response function of two-particle operators in general. 
As an example, in Fig.~\ref{F:ddr} we plot the strength function of the $0^+$ component of the double dipole resonance. 
We notice that the SRPA0 strength function is very close to the unperturbed one (HF) and differs significantly from the SRPA result.

\begin{figure}[!]\centering
\includegraphics[angle=-90,width=8cm]{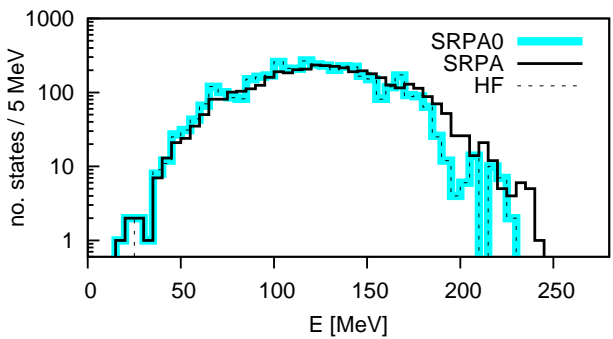} 
\caption{%
For the $0^+$ response of $^{16}$O, calculated within a single-particle basis of 7 oscillator shells: 
Number of SRPA and of SRPA0 (diagonal approximation, Eq.~(\ref{Ea22}) eigenstates per 5~MeV excitation energy --- in logarithmic scale. The corresponding density of unperturbed (HF) $ph$ and $2p2h$ states is also shown.  
\label{F:eigen05}}
\end{figure}

In Fig.~\ref{F:eigen05} we show the density of eigenstates (number of states per 5~MeV excitation energy) when solving the SRPA problem with and without the diagonal approximation, as well as the density of unperturbed $ph$ and $2p2h$ states (HF). 
One immediately notices that for the most part the SRPA0 density of states is very similar to the unperturbed one.

Finally, below about 40~MeV, where the giant resonance lies, the SRPA and SRPA0 densities practically coincide. This is not surprising, given the very small number of eigenstates present there, but it is not always the case. 
We note, for example, that in the case of the $2^+$ response of $^{48}$Ca, 
where there are several unperturbed $2p2h$ states at low energies, the SRPA0 density of states follows mostly the HF one 
in that energy region. 
The single-particle strength functions, however, are almost identical (Fig.~\ref{F:full}a)  
and the diagonal approximation remains well justified. 


\subsection{Downward shift of resonances} 

\label{S:shift} 

In Figs.~\ref{F:toy} and \ref{F:full} we notice that the GR lies lower in SRPA(0) than in RPA. 
This is a rather general result ---  see also Ref.~\cite{PaR2009}. 
It is a general feature of the coupling to $2p2h$ excitations~\cite{Wam1988,LAC2004} (and not particular to nuclei~\cite{GaC2009}), 
and shows why traditional effective interactions cannot be used in SRPA. 
The self-energy corrections responsible for the modification of the $ph$ energies and the lowering of the resonance energies (see also Sec.~\ref{S:fragm}) 
are already parameterized in the interaction, 
so that realistic results are obtained already at the RPA level. 
Employing such an interaction in SRPA will result in double counting of those effects. 

Ths problem is circumvented in practical applications in the literature through the use of 
realistic single-particle energies and subtracting procedures, 
which remove the real part of the self energy, responsible for the shift.


\subsection{Fragmentation in SPRA} 

\label{S:fragm} 


As a microscopic theory of collisional damping, SRPA has been used extensively 
to describe the spreading width and strength fragmentation and quenching of collective excitations.  
Let us examine our results in this context. 

No spreading width is observed in Fig.~\ref{F:toy}a, as there are no configurations available in the vicinity of the resonance. 
But when we look at Fig.~\ref{F:toy}b, we realize that there is practically a continuum of weak SRPA or SRPA0 eigenstates ($N=4148$) extending to high energies. Most are predominantly of $2p2h$ nature and their contribution to the single-particle strength is rather small. Nevertheless, they carry a non-negligible percentage of the total strength and provide a mechanism of strength quenching for the giant resonance (by contrast, there are only $N_1=14$ RPA eigenstates). We note, in particular, that: 
(i) As expected, the total energy-weighted strength $m_1$ is found to be practically the same in all three cases (RPA, SRPA0, SRPA), and the total strength $m_0$ is found just about $3\%$ larger in SRPA or SRPA0 than in RPA, but 
(ii) the strongest ISM peak appears at 21.36~MeV in the case of RPA, 16.50~MeV in SRPA0, and 16.25~MeV in SRPA and (iii) the six lowest eigenstates carry a total strength of 313.4~fm$^4$ in RPA, which is almost all the RPA strength (they lie at 21 to 47~MeV excitation energy), 304.5~fm$^4$ in SRPA0 (16 to 34~MeV) and 304.0~fm$^4$ in SRPA (16 to 34~MeV), i.e., about $3\%$ less in SRPA(0) than in RPA; finally, (iv) the width of the ISM strength functions below 40~MeV is 3.86~MeV in RPA and 3.73~MeV in SRPA, while the widths over the whole spectrum are 4.65 and 18.94~MeV respectively. 

\begin{figure}[!]\centering
\includegraphics[angle=-90,width=8cm]{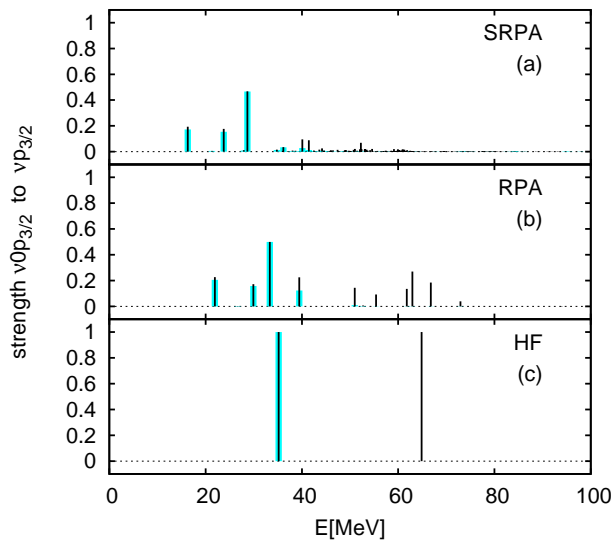} 
\caption{%
Fragmentation and shift of $ph$ states --- $0^+$ response of $^{16}$O in a space of 7 shells. 
Thin dark bars show how the spectroscopic strength $S_{ph}(E)$, Eq.~(\ref{Esph}), of the $ph$ configurations $|(\nu p_{3/2})(\nu 0p_{3/2})^{-1};0^{+}\rangle$ (contributing to the monopoole strength) is distributed within {(a)} SRPA, {(b)} RPA and {(c)} HF. 
Thicker, pale (cyan) bars denote the distribution of $|(\nu 1p_{3/2})(\nu 0p_{3/2})^{-1};0^+\rangle$ (one particle shell only). 
\label{F:PHfragLin}}
\end{figure}

The coupling with $2p2h$ configurations affects not only the collective excitations, but at the same time the single-particle states. 
Within the present formalism one can demonstrate this through the strength of the $ph$ configurations. 
In Fig.~\ref{F:PHfragLin}, for example, we plot the HF, RPA and SRPA strength distribution 
$S_{ph}(E)$, Eq.~(\ref{Esph}), of the $ph$ configurations $|(\nu p_{3/2})(\nu 0p_{3/2})^{-1};0^{+}\rangle$, contributing to the $0^+$ strength of 
Fig.~\ref{F:toy}. 
In HF, these are well-defined transitions at energies equal to $e_p-e_h$ with strength 1. 
In RPA the $ph$ transitions appear fragmented. There is a cluster of configurations at energies around 30~MeV, which can be attributed to the $2\hbar\omega$ transition 
$|(\nu 1p_{3/2})(\nu 0p_{3/2})^{-1};0^+\rangle$. 
A second cluster of states is visible around 60~MeV roughly corresponding to $4\hbar\omega$ configurations, and so on. 
In SRPA the strength distribution appears even more fragmented, as well as shifted to lower energies. The shift reflects an effective compression of the single-particle spectrum (a modification of the nucleon effective mass), with respect to HF, and leads to the downward shift of the collective states, discussed above. 

\begin{figure}[!]\centering
\includegraphics[angle=-90,width=8cm]{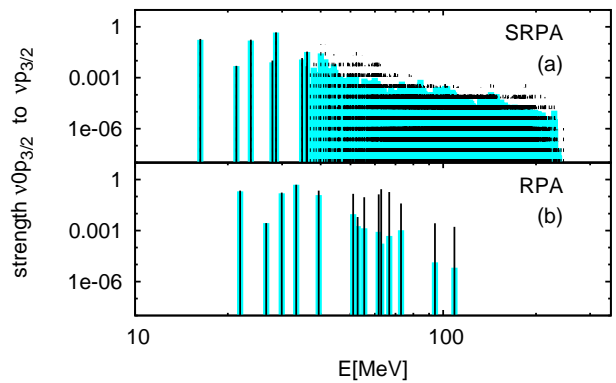} 
\caption{%
Same as Fig.~\ref{F:PHfragLin} without HF, both axes in logarithmic scale. 
\label{F:PHfragLog}}
\end{figure}

Although we can visually identify the different shells contributing to the strength distribution, 
and observe an energetic shift, the total $m_0$ and $m_1$ --- and thus the centroid --- of any given $ph$ configuration is the same in RPA and SRPA. 
Nevertheless, in RPA we find 
only 1\% of the 
$|(\nu 1p_{3/2})(\nu 0p_{3/2})^{-1};0^+\rangle$ 
$m_0$ strength above 40~MeV. In SRPA the strength in the same region is 14\%. 
When we look at the distributions in logarithmic scale, Fig.~\ref{F:PHfragLog}, we realize that the 
$ph$ strength distributions span the whole space available, 
reflecting the extended spectroscopic functions of the particle and hole states. 


\subsection{Truncation procedures} 

\label{S:trunc} 

\begin{figure}[!]\centering
\includegraphics[angle=-90,width=8.5cm]{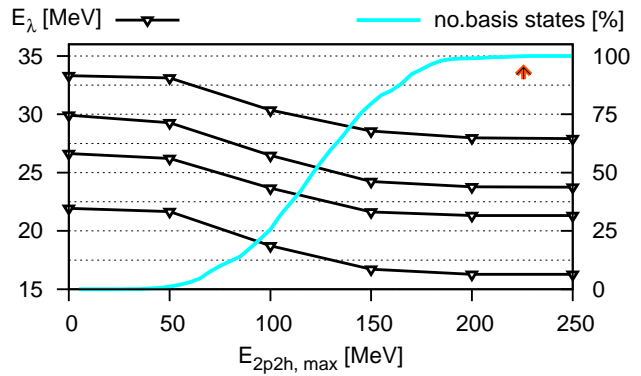} 
\caption{%
For the $0^+$ response of $^{16}$O, calculated within a single-particle basis of 7 oscillator shells, but imposing an energy cutoff on the $2p2h$ configurations taken into account, $E_{\mathrm{ph, max}}$: 
Energies of the four lowest eigenstates (axis on the left) and percentage of basis states taken into account 
(axis on the right), vs $E_{\mathrm{ph, max}}$. 
The arrow indicates the point at which all available states are used (maximal $2p2h$ energy). 
\label{F:Eph2Max}}
\end{figure}

As is obvious from Figs.~\ref{F:toy} and \ref{F:eigen05}, the majority of basis states (unperturbed $ph$ and $2p2h$ states) lies at high energies relative to the giant resonance. Figure~\ref{F:Eph2Max} (see caption for details) illustrates what would happen if we truncated the model space by setting an upper cutoff $E_{2p2h,\mathrm{max}}$ to the $2p2h$ energies taken into account. Had we set, e.g., $E_{2p2h,\mathrm{max}}=100$~MeV, we would have excluded 75\% of the configurations available in this particular space and the eigenenergies would have been at least 2~MeV higher than their converged (with respect to $E_{2p2h,\mathrm{max}}$) values. Inclusion of the energetically lowest 50\% of configurations would still not warrant good convergence. 

We found that the convergence of the eigenenergies and the saturation of the model space follow the same pattern as demonstrated in Fig.~\ref{F:Eph2Max}, regardless of model-space size and type of response. 

Finally, we should note that most finite $\mathcal{A}_{12}$ elements are small. 
The distribution of their values is always cusp shaped around zero 
and a vast majority of them have amplitutudes no larger than 5\% the value of the strongest element. It would be computationally economical to neglect those elements. 
We found, however, that such a procedure influences the position of the resonances as well as the shape of the strength distribution and cannot be blindly trusted. 

\subsection{Consistency and stability issues} 

\label{S:cons} 

\begin{figure}[!]\centering
\includegraphics[angle=-90,width=8cm]{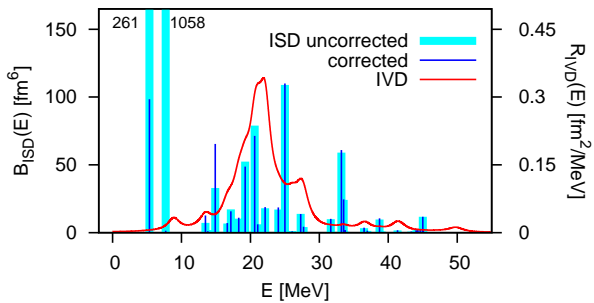} 
\caption{%
Spurious admixtures in the SRPA dipole spectrum. Dipole strength distributions of $^{16}$O using $V_{\mathrm{UCOM}}$ and a basis with $e_{\max}=12$, $\ell_{\max}=10$. The scale on the left corresponds to the IS strength and the scale on the right to the folded IV strength. The IS strength is shown for the usual IS dipole transition operator and for its uncorrected form --- see text. 
\label{F:spurious}}
\end{figure}

The self-consistent RPA that we have used produces a spurious state at practically zero energy, 
and leaves the rest of the spectrum uncontaminated~\cite{PPH2006}, as expected. 
The SRPA dipole spectrum, however, may contain spurious admixture, as discussed in Sec.~\ref{S:cons1}. 
In order to quantify this problem, we have examined the IS dipole response. 
We found that relatively strong spurious states appear mostly at about 5 to 8~MeV. 
Using a transition operator of the usual radial form ($\propto r^3 - \frac{5}{3}\langle r^2 \rangle r$) and its uncorrected form ($\propto r^3$), we found that only the lowest part of the dipole spectrum is strongly affected by the choice of operator, while there are no significant contaminations in the spectrum around and beyond the IV GDR peak. An example is shown in Fig.~\ref{F:spurious}.  

Let us note, that including a small percentage of $2p2h$ states would not shift the spurious state far away from zero, but for large spaces such as those used here the effect is noticeable. 
As we couple the $ph$ states to more and more $2p2h$ configurations, the spurious RPA state moves away from zero and may occur at imaginary, or even negative energies (or, equivalently, positive energies, but assuming negative norm --- this is the case in Fig.~\ref{F:spurious}). 
It may also appear fragmented, as spurious admixtures enter the rest of the spectrum, or as additional eigenstates occur in its proximity. 

As anticipated in Sec.~\ref{S:cons1}, another problem with the HF-based SRPA is  the onset of instabilities. We find, in particular, that 
low-lying states ($0\hbar\omega$ $2^+$ states and the collective $3^-$ excitation) appear at imaginary or negative energies. 
This is the case, for example, in Fig.~\ref{F:full}a: the norm of the collective quadrupole state at $4.4$~MeV is, in fact, negative; the positive-norm counterpart appears at $-4.4$~MeV. 
As discussed previously, the problem seems to be the inadequate treatment of ground-state correlations.

\begin{figure*}[!]\centering
\includegraphics[angle=-90,width=17cm]{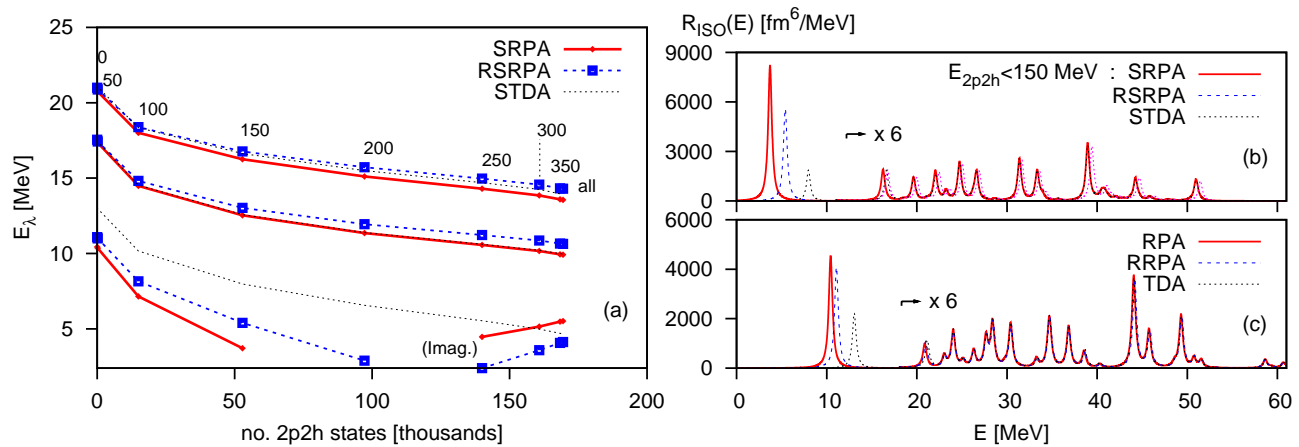} 
\caption{%
Exploring the influence of ground state correlations: Isoscalar $3^-$ response of $^{16}$O ($\epsilon_{\max}=14$, $\ell_{\max}=10$) within RPA or SRPA and ``renormalized" RPA or SRPA, (RRPA, RSRPA --- see text), as well as TDA and STDA (SRPA with $B=0$). 
{(a)}: Evolution of the three lowest (closest to zero) eigenvalues as we increase the number of $2p2h$ configurations considered, starting from zero and up to all $2p2h$ states available in the single-particle space, $N_2$. The numbers assigned to the various points indicate the corresponding $2p2h-$energy cutoff imposed, $E_{2p2h,\max}$, in MeV. The energy of the lowest eigenstate becomes imaginary for larger $N_2$ in (R)SRPA; then its amplitude is indicated. Note that the STDA and SRPA results for the second state almost coincide.  
{\em (b)}: Strength function for $E_{2p2h,\max}=150$~MeV in SRPA, RSRPA, and STDA ($\Gamma = 2$~MeV). 
{\em (c)}: Strength function in RPA, RRPA, and TDA. 
\label{F:GSC}}
\end{figure*}

Obviously, SRPA is not appropriate for describing low-lying states. The question is then whether it still gives meaningful results for giant resonances. 
We have tested the sensitivity of our results to GSC with the help of the renormalized SRPA devised in Sec.~\ref{S:rsrpa} and of STDA (SRPA setting $B=0$) and, in first order, the corresponding RRPA and TDA. 
Examples are shown in Fig.~\ref{F:GSC} for the octupole response of $^{16}$O and Table~\ref{T:GSC} for the quadrupole response of $^{48}$Ca. In all cases the diagonal approximation has been employed.  
%
\begin{table}[t]
\centering 
\begin{ruledtabular} 
\begin{tabular}{l|ccc|ccc} 
       &  $E_1$  &  $E_2$  &  $E_{\mathrm{GQR}}$  &  $B(E_1)$  &  $B(E_2)$  &  $B(\mathrm{GQR})$  \\ 
\hline 
RPA    &  2.19  &  8.12  &  27.22              &   450.05   &    79.18   &   915.2            \\  
RRPA   &  2.42  &  8.16  &  27.41              &   373.09   &    69.36   &   892.2            \\  
TDA    &  2.61  &  8.39  &  27.42              &   127.05   &    46.06   &   813.1            \\  
\hline 
SRPA  & -4.44  &$i\times$0.803 
                           &  19.51              &   223.18   &     ---     &  1021.3            \\  
RSRPA & -3.14  &  1.34  &  20.18              &   161.95   &    22.25   &   991.3            \\  
STDA  & -4.26  &  0.46  &  19.72              &   182.28   &    41.14   &   831.1            \\  
\end{tabular}  
\end{ruledtabular}  
\caption{%
Isoscalar $2^+$ spectrum of $^{48}$Ca calculated in a basis of 9 oscillator shells: 
Energy $E$ (in MeV) and strength $B$ (in fm$^4$) of the two main low-lying states and of the giant resonance peak (GQR) 
within RPA and SRPA, ``renormalized" RPA and SRPA (RRPA, RSRPA --- see text), and TDA and STDA (SRPA with $B=0$). 
(Whenever the GQR is split into two or three major peaks, the centroid and total strength is given.) 
\label{T:GSC}
} 
\end{table} 

The general trend can be described as follows: 
We begin with minimal GSC, in STDA. Inclusion of the $B$ matrix and the backward amplitudes (SRPA) pushes the solutions to somewhat lower energies and in general affects their strength. 
The downward shift, which can lead to imaginary solutions, 
was interpreted recently in the framework of random-matrix theory~\cite{BBW2009}: 
The matrix $B$ couples the positive-and negative-energy branches of the (S)RPA solutions and causes an attraction between them. Strong enough coupling leads to a merging of the two branches at zero energy 
and eventually to imaginary or complex solutions. 
Finally, we find that subsequent renormalization of the matrix elements (RSRPA) shifts the solutions back to higher energies, i.e., slightly closer to the RPA solutions, 
plausibly because of the weaker $\mathrm{A}_{12}$ couplings.

The effect of renormalization appears rather small in the case of giant resonances and higher-lying solutions in general, 
but this is not the case for the low-lying quadrupole and octupole states: their energy shifts in RSRPA, relative to RPA, are noticably more moderate than in SRPA. 
The STDA results confirm in general that the lower-lying states are more sensitive to GSC than the higher ones. 
We note that, since STDA is a Hermitean problem, it will eliminate the imaginary solutions of SRPA by construction --- see, e.g., Fig.~\ref{F:GSC}a. As we observe in Table~\ref{T:GSC}, however, the same does not hold for the negative solutions. 
It is thus confirmed that 
the strong resonance shifts (with respect to RPA) in this case are induced by the $\mathcal{A}_{12}$ couplings, as discussed in Sec.~\ref{S:cons1}. 

In most cases the RSRPA solutions lie higher than the STDA, i.e., renormalization has a stronger effect on the energies, with the notable exception of the giant monopole resonance (not shown), where the backward amplitudes seem more relevant. 
 
Finally, regarding our first-order results, renormalization affects the RPA results very weakly. 
The same conclusion was reached in Ref.~\cite{PRP2007}, where a more consistent RRPA was applied. 
The backward amplitudes, though, missing in TDA, are found important for the description of low-lying states.  

The present results suggest that giant resonances (at least for $J>0$) in SRPA are only moderately sensitive to the treatment of GSC. 
Of course, a comprehensive inclusion of GSC in the SRPA formalism would involve filling up the 
$B-$sections of the SRPA matrix with many more finite elements. It cannot be predicted how strongly those could affect the results. In the applications shown in Ref.~\cite{DNS1990}, the so-called extended SRPA including correlations did not produce strong corrections to the SRPA strength functions, especially after renormalizing the ground state~\cite{VWV1992,MKd1994}. 


\section{Conclusions} 

\label{S:concl} 

The present work constitutes a feasibility and justification study 
of large-scale SRPA calculations. 
The motivation was the prospect of studying nuclear collective excitations 
using SRPA and unitarily transformed nuclear Hamiltonians, which 
have not been fitted to first-order RPA results. 
The discussion, however, has been kept general. 
We showed how the large model spaces of SRPA can be treated 
and discussed the salient features of the solutions, including the 
energetic shift relative to the RPA solutions and the fragmentation of strength. 
The diagonal approximation was found reliable for soft interactions 
and for single-particle strength distributions. 

We found that low-lying states become unstable in SRPA, due to an inadequate treatment 
of ground-state correlations. 
Nevertheless, giant resonances and higher-lying solutions in general do not appear sensitive 
to ground-state correlations. 
We have thus concluded that SRPA 
can be applied in the giant-resonance region with reasonable confidence. 

SRPA is primarily the theory of collisional damping. 
A more comprehensive method to study nuclear collective states 
should consider also coupling to collective low-lying phonons~\cite{BBB1983,Sol1992}. 
It is not easy to tell at this point how strongly those could enhance or 
(partly) cancel the effect of coupling to the large amount of $2p2h$ states 
considered in SRPA. In Ref.~\cite{LAC2004} it was shown that the resonance 
shifts due to the $2p2h$ states are stronger 
than those due to collective phonons, although both mechanisms are expected 
to contribute to the damping width.  

Notwithstanding some particular shortcomings of the present SRPA formalism, 
this work paves the way for systematic studies of giant resonances 
using finite-range interactions and extended RPA theories without adjustable 
parameters and arbitrary truncations of the model space.

\begin{acknowledgments} 
Discussions with Vladimir Ponomarev and Jochen Wambach are gratefully acknowledged. 
This work was supported by the Deutsche Forschungsgemeinschaft 
through SFB 634, by the Helmholtz International Center for FAIR 
within the framework of the LOEWE program launched by the state of Hesse, 
and by the BMBF Verbundforschung (Contract 06DA90401).  
\end{acknowledgments} 


%
%



\end{document}